# Pulsar Timing Observations with Haoping Radio Telescope

Jin-Tao Luo[1,2], Yu-Ping Gao[1,3], Ting-Gao Yang[1], Cheng-Shi Zhao[1], Ming-Lei Tong[1], Yong-Nan Rao[1,2], Yi-Feng Li[1,2], Bian Li[1], Xing-Zhi Zhu[1], Hai-Hua Qiao[1] and Xiao-Chun Lu[1,2]

[1] National Time Service Center, Chinese Academy of Sciences, Xi'an 710600, China; *jluo@ntsc.ac.cn*

[2] Key Lab. of Precision Navigation & Timing Technology, Chinese Academy of Sciences, Xi'an 710600, China

[3] University of Chinese Academy of Sciences, Beijing 100049, PR China

**Abstract** We report pulsar timing observations carried out in L-band with NTSC's 40-meter Haoping Radio Telescope (HRT), which was constructed in 2014. These observations were carried out using the pulsar machine we developed. Timing observations toward millisecond pulsar J0437-4715 obtains a timing residual (r.m.s) of 397ns in the time span of 284 days. And our observations successfully detected Crab pulsar's glitch that happened on July 23rd, 2019.

## 1 INTRODUCTION

The Haoping 40-meter radio telescope (HRT), operated by National Time Service Center (NTSC), is a new radio astronomy facility in China. HRT is located in the Qinling mountains, ∼ 100 km east of the city of Xi'an.

As a national unit for scientific research, NTSC is an institute dedicated to carrying out the professional research on time and frequency sciences. Among NTSC's research areas, the pulsar time plays an important role. At NTSC the pulsar timescale, and the assemble pulsar timescale, and the application of pulsar timescale have been well studied (Yang 2003; Yin et al. 2016; Tong el al. 2017). These studies are mainly on the theory and algorithms, using the data from PTAs (Pulsar Timing Arrays), such as Parkes Pulsar Timing Array (PPTA) in Australia, which are abroad of China. With the construction of HRT and the development of a pulsar machine, NTSC started pulsar timing observations from the year of 2018.

This paper will present an introduction to the 40-meter Haoping radio telescope and the pulsar machine developed for this telescope. The pulsar timing observations taken with this system and the latest timing results will be reported as well, including millisecond pulsar timing and glitch monitoring.

## 2  PULSAR TIMING OBSERVING SYSTEM

**Radio Telescope**

The Haoping radio telescope was constructed in 2014. As shown in figure 1, the site is located deep in the mountains. Benefits from the isolation from cities and the protection of surrounding little hills, the site is of a very good radio environment. However the nearby hills do not cause serious blockage to the telescope. Objects with elevation angle larger than 7° are visible to the telescope.

As shown in figure 2 the telescope is a Cassegrain antenna system. The main reflector, which consists of full solid panels, is installed on an altitude-azimuth mount. The surface accuracies (r.m.s) of the main and secondary reflectors are $\leq$ 0.6 mm and $\leq$ 0.15 mm, respectively. Table 1 lists some key specifications and performances of the telescope.

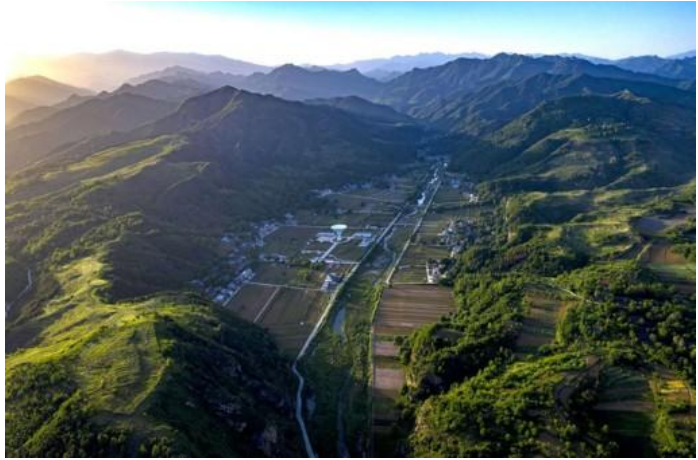

Fig. 1: The bird view of Haoping 40-meter Radio Telescope.

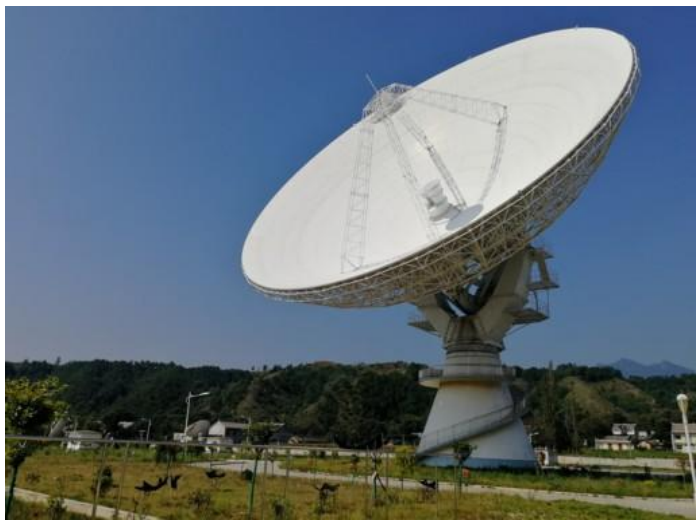

Fig. 2: The close view of Haoping 40-meter Radio Telescope.

Table 1: Some Key Specifications and Performances of HRT.

| Spec/Para | Value |
|---|---|
| AZ range | ± 320° |
| EL range | 5° ~ 92° |
| Max. AZ Speed | 1.5°/sec |
| Max. EL Speed | 1°/sec |
| Min. Speed | 0.0005°/sec |
| Pointing Accuracy | ≤1/10 Beam Width |

**Receivers**

HRT is equipped with L-band and S-band receivers. Horns of these receivers can be seen in figure 2. The L-band receiver, which is used to carry out pulsar observations, covers the frequency range of 1.1 ~ 1.75GHz. Before November 2019, we conducted single-polarisation observations with HRT as only the RCP (Right Circular Polarisation) was received. Since then, with the installation of the LCP (Left Circular Polarisation) receiving system, both two polarisations are available. As shown in figure 3, the RCP and LCP RF (Radio Frequency) signals, are amplified using the room-temperature LNAs (Low Noise Amplifiers) in the cabin and then transferred to the control room via the fiber using the RF-over-Fiber technology.

At the control room the RF signals are down converted using the LO (Local Oscillator) frequency of 1GHz, producing IF (Intermediate Frequency) signals with the bandwidth of 800MHz. The properly-amplified IF signals are then sent to the pulsar machine for further processes.

The system temperature of the L-band receiving system is about 100K.

**Pulsar Machine**

IF signals from the receiving system are processed in the pulsar machine, such a backend development uses the FPGA+GPU architecture at HRT. Table 2 lists some key specifications of the machine .

As shown in figure 3 IF signals are first digitised using the high speed ADCs (Analog-to-Digital Converters), whose sampling clock rates can be adjusted to meet the observing requirement. Digital IF signals then get processed on the FPGA (Field Programmable Gate Array) platform, which is based on the ROACH2 board developed by University of California (UC), Berkeley in the project of CASPER (Hickish et al. 2016). On the FPGA the digital IF signals first get divided into sub-channels using filter banks. In the coherent dedispersion mode the baseband data of these channels are directly sent out. In the incoherent dedispersion mode the calculation of four Stokes parameters, and the following process of accumulation which aims to reduce the data rate, are applied to each sub-channel.

The output of the FPGA is transferred to a 8-node GPU (Graphics Processing Unit) cluster via one 10GbE (10 Giga bit Ethernet) interface in the incoherent dedispersion mode, and via eight 10GbE interfaces in the coherent dedispersion mode. In the incoherent mode the output data is processed on one node of the cluster. Depends on the observation type, the machine stores the search mode raw data directly or fold the data using a given ephemeris file and then store the sub-integration profiles onto the storage. The stored data is in the PSRFITS format. Similar operations are implemented on the 8 nodes of the cluster in the coherent mode, while each node processes one eighth of the total observing band. The pulsar machine is controlled using a computer through the 1GbE (1 Giga bit Ethernet) links.

Table 2: Specifications of Pulsar Machine at HRT.

| Incoherent Dedispersion Mode | |
|---|---|
| Obs. Mode | Fold, Search |
| Maxi. Bandwidth | 2 GHz |
| No. of Channels | 8192, 4096, 2048, 1024, 512, 256, 128, 64 |
| Highest Time Resolution | 10.24 $\mu$s |
| Coherent Dedispersion Mode | |
| Obs. Mode | Fold, Search |
| Maxi. Bandwidth | 1.6 GHz |
| No. of Channels | 4096, 2048, 1024, 512, 256, 128, 64 |
| Highest Time Resolution | < 10.24 $\mu$s |

The 10MHz signal from the on-site Cesium atomic clock is used to generate input clocks for ADCs. And this clock's 1PPS (Pulse Per Second) signal is used to trigger the pulsar machine. Additionally this atomic clock is synchronized to UTC(NTSC) with an error less than 2ns using the satellite CV (Common View) technique. These ensure precise and stable time and frequency signals, together with an accurate on-site time reference for pulsar timing observations.

PSR J1939+2134, a millisecond pulsar with a period of ∼ 1.56 ms, is used to test the coherent dedispersion functionality. The coherent dedispersion technology can recover the true pulse shape of this pulsar (McMahon 2008). Figure 4 shows the integrated profile of PSR J1939+2134 obtained with our pulsar machine in coherent dedispersion mode. As shown in this figure the fine structure of this pulsar's main pulse, around phase 0.75, was recovered.

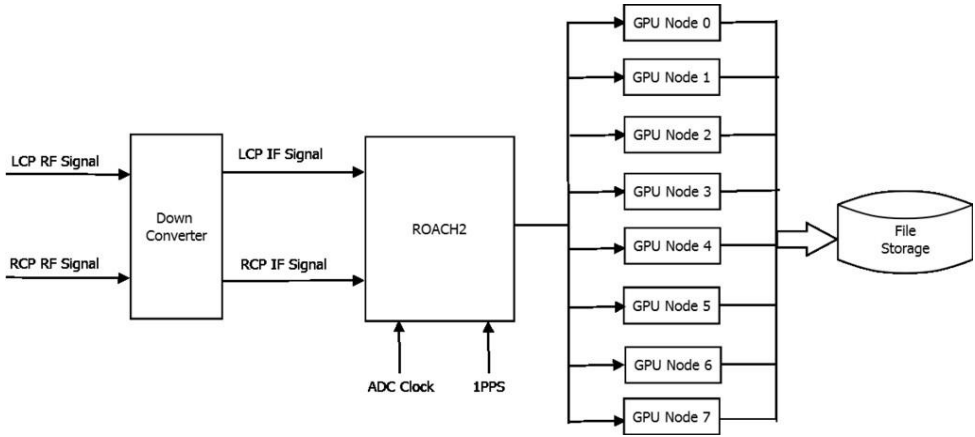

Fig. 3: Diagram of the Pulsar Timing System.

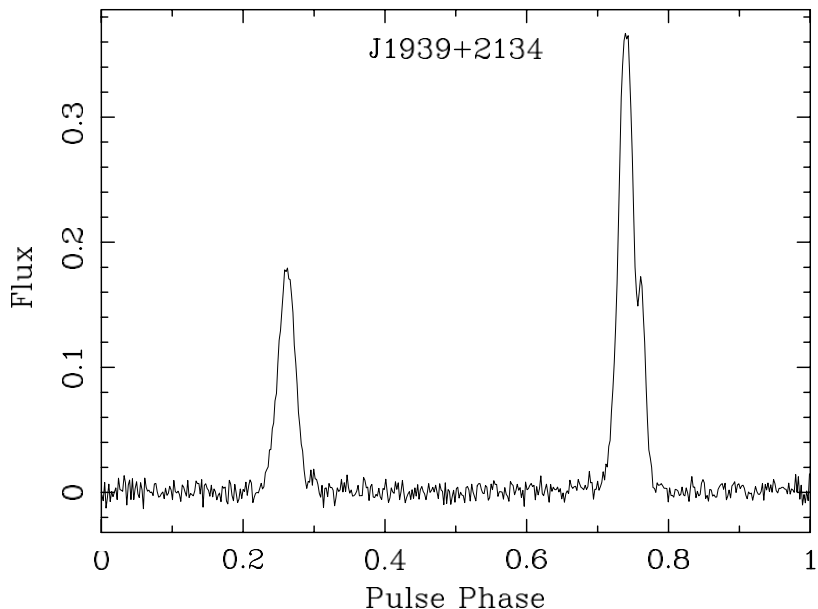

Fig. 4: The integrated profile of PSR J1939+2134, the flux density is plotted in arbitrary units.

Table 3: Brief Parameters and Observing Configurations of Millisecond Pulsars.

| Pulsar Name | RaJ(h:m:s) | DecJ(d:m:s) | Period(ms) | DM($cm^{-3}$ pc) | Obs. Cadence | Obs. Length |
|---|---|---|---|---|---|---|
| J0437−4715 | 04:37:15.90 | −47:15:09.11 | 5.757 | 2.64476 | ∼3 days | > 40min |
| J1713+0747 | 17:13:49.53 | +07:47:37.48 | 4.570 | 15.917 | ∼3 weeks | > 40min |
| J1939+2134 | 19:39:38.56 | +21:34:59.13 | 1.558 | 71.0237 | ∼1 month | > 40min |
| J1909−3744 | 19:09:47.43 | −37:44:14.46 | 2.947 | 10.3932 | ∼1 month | > 40min |
| J1022+1001 | 10:22:57.99 | +10:01:52.77 | 1.645 | 10.2521 | ∼1 month | > 40min |
| J1643−1224 | 16:43:38.16 | −12:24:58.68 | 4622 | 62.4143 | ∼1 month | > 40min |
| J2145−0750 | 21:45:50.46 | −07:50:18.51 | 1.605 | 8.99761 | ∼1 month | > 40min |

## 3 MILLISECOND PULSAR TIMING

**Observation**

The long-term millisecond pulsar timing observing program started from the end of August 2018. Several millisecond pulsars are being regularly observed with cadences range from ∼ 3 days to 1 month. The brief parameters and observing configurations of these millisecond pulsars are listed with table 3.

Before August 2019, observations were carried out in the incoherent dedispersion mode, and since then observations were taken using the coherent dedispersion mode. In both modes 1024-channel mode were used, together with 1024 phase bins. And the length of the sub-integration is set to 10 seconds, in order to keep a good time resolution for the detection and mitigation of time-domain burst interferences in the data process. For observations in the incoherent mode, the time resolution is set to 10.24 $\mu$s.

These observations use the ephemeris from PSRCAT[1] (Manchester et al. 2005) and NANOGrav (Arzoumanian el al. 2018) and PPTA[2] (Reardon et al. 2015).

---
[1] http://www.atnf.csiro.au/research/pulsar/psrcat
[2] https://data.csiro.au/collections/#collection/CIcsiro:13081v3

**Data Reduction and Analysis**

The data reduction and analysis is implemented using the pulsar software PSRCHIVE (Hotan et al. 2004). The raw data, which stores sub-integration profiles, first gets updated with ephemeris released from PPTA and(or) NANOGrav using the program PAM. Then the program PAZI is used to do manual RFI mitigations. No calibration is applied as HRT has not been equipped with flux and polarisation calibration capability.

TOAs (Time Of Arrivals) are generated with the program of PAT, using profiles released from PPTA and NANOGrav as the template. Timing analysis is then carried out using the software TEMPO2 (Hobbs et al. 2006).

**Results**

We take J0437−4715 as the example of our millisecond pulsar observations. Its integrated profile is shown in figure 5. With an integration time of ∼ 75 minutes, the profile obtained with our system shows an SNR of ∼ 640.

Figure 6 shows the timing residual of PSR J0437−4715. In the time span of 284 days, the obtained timing residual (r.m.s) is 397ns. As can be seen in this figure, the distribution of TOA dots are not random which suggests there are systems errors need to be removed. The lack of polarization calibration is a likely cause of these errors.

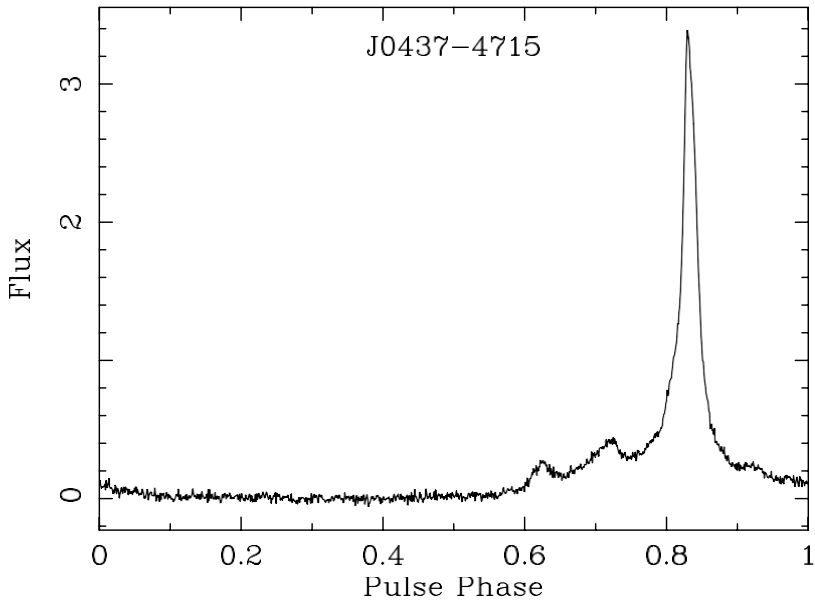

Fig. 5: The integrated profile of PSR J0437−4715, the flux density is plotted in arbitrary units.

## 4 GLITCH MONITORING

**Observation**

A few pulsars which have glitch records are being monitored via timing observations at HRT. Configurations of these observations are the same as introduced in section 3.1. With these observations we managed to detect the Crab pulsar's glitch event that happened on July 23rd, 2019. In timing observations towards Crab pulsar (PSR J0534+2200) the ephemeris from PSRCAT is used and the observing length varies from 20 minutes to 2 hours.

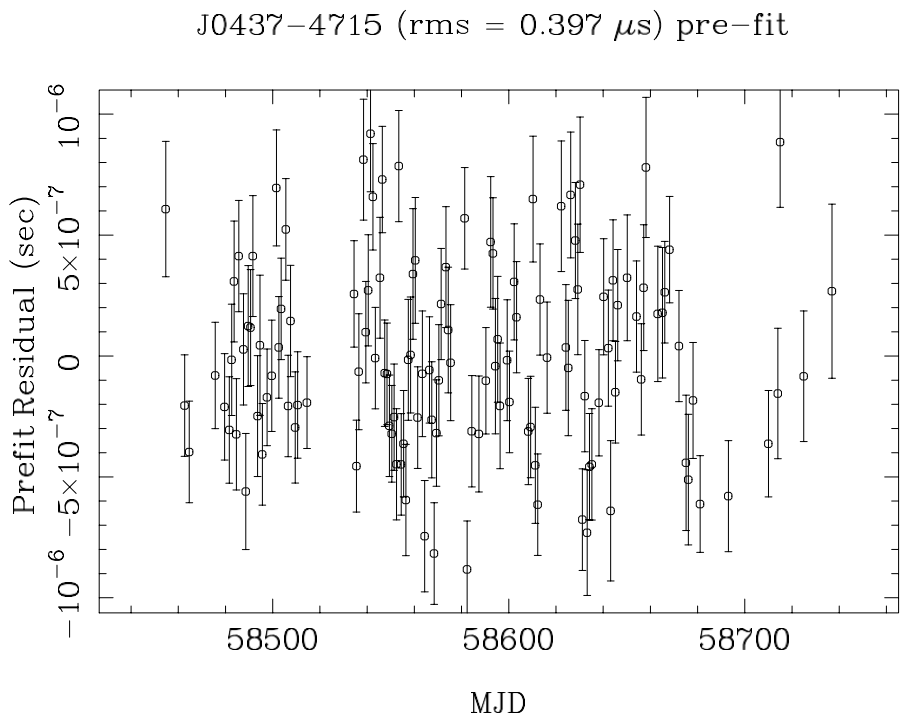

Fig. 6: Timing residual (r.m.s) of PSR J0437−4715.

**Data Reduction and Analysis**

The process is the same as that introduced in section 3.2. For Crab pulsar, the monthly ephemeris released by Jodrel Bank Observatory[3] (Lyne et al. 1993) is used to update the data obtained with HRT. A high-quality profile obtained from our observations is used as the template.

According to the glitch catalog (Espinoza et al. 2011)[4], in the course of our observations Crab pulsar had two glitch event which are reported to happen on 18 December 2018 and 23 July 2019. In this paper we report the analysis on the latter one. TOAs from January to July in 2019 (before the event) are fitted to determine parameters before the glitch. This pre-glitch ephemeris is then used in the timing process on all the TOAs from January 2019, including TOAs before and after the July glitch event.

**Results**

Figure 7 shows an integrated profile of Crab pulsar obtained from our observations. The integration time is ∼ 47 minutes, and the profile's SNR is ∼ 44. This profile clearly shows Crab pulsar's main pulse and interpulse, around phase 0.3 and 0.7 respectively. And in this figure the precursor which is ∼ 0.1 phase before the main pulse (Lyne et al. 2013) can be seen as well.

The timing residual relative to the pre-glitch timing model is shown in figure 8. The corresponding observations cover the period from MJD 58534 to MJD 58825. This residual shows an offset-and-return structure, which is typical in Crab pulsar's glitch timing residulas (Wang et al. 2001). Using the last pre-glitch data and the first post-glitch data in our observations, the estimated glitch epoch is MJD 58684.621 with the uncertainty of 8.584 days. The glitch epoch reported from the glitch catalog is MJD 58687.59. Our detection coincides with the report. The post-glitch TOAs show obvious offsets from the pre-glitch model and begin to return to this model from ∼ MJD 58750.

---
[3] http://www.jb.man.ac.uk/~pulsar/crab.html
[4] http://www.jb.man.ac.uk/pulsar/glitches/gTable.html

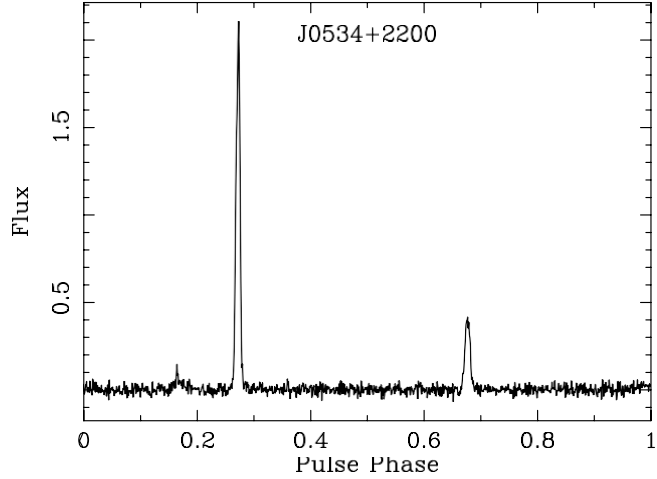

Fig. 7: The integrated profile of Crab pulsar, the flux density is plotted in arbitrary units.

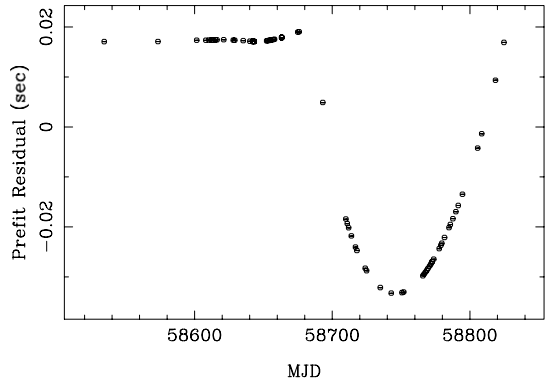

Fig. 8: The timing residual of Crab pulsar from MJD 58534 to MJD 58825.

## 5 SUMMARY

We have developed a pulsar timing observing system at our 40-meter radio telescope. Using this system several millisecond pulsars and glitch pulsars are being observed. Timing observations toward millisecond pulsar J0437−4715 obtained a timing residual (r.m.s) of 397ns in the time span of 284 days. And timing observations toward Crab pulsar, J0534+2200, has successfully detected this pulsar's glitch in July, 2019. High-quality profiles of these two pulsars obtained with our system, together with the good timing residual on millisecond pulsar J0437−4715 and the detection of Crab pulsar's glitch, indicate that this system has a pretty good receiving sensitivity and system stability for pulsar timing observations.

After installing the receiving system for the LCP signal, we are working on the calibration capability of our pulsar timing system. This effort is expected to improve the timing result of millisecond pulsars like J0437−4715.

**Acknowledgements** This work was supported by the Chinese Academy of Sciences Hundred Talents Program ( Technological excellence, Y650YC1201), the National Natural Science Foundation of China ( Grant Nos. U1931128, 11973046, 91736207, U1831130, 11903038, 11873050, 11873049). Shaanxi province National Natural Science Foundation (Grant No. 2019JM-455), and the program of Youth Innovation Promotion Association CAS (2017450).